\documentclass[10pt]{article}
\usepackage{graphicx}

\usepackage{epsfig}
\usepackage{latexsym}
\usepackage{amsmath}
\begin{document}

\title{A biased   view of a few possible components \\when reflecting on    the present decade financial and economic crisis }
\author{  M. Ausloos$^{}$\footnote{ {\it previously at} GRAPES$@$SUPRATECS, University of Li\`ege,   B-4000 Li\`ege, Euroland; \;\;\;\;\;\;\;\;$e$-$mail$ $address$:
marcel.ausloos@ulg.ac.be} \\  $^{}$ Beauvallon, r. de la Belle Jardiniere, 483/0021, \\
B-4031 Liege, Wallonia-Brussels Federation, Belgium}

 \date{Oct. 02, 2012}
\maketitle
\vskip 0.5truecm

\begin{abstract}
Is the present economic and financial crisis similar to some previous one? It would be so nice to prove that  universality laws exist for predicting such rare events under a minimum set of realistic hypotheses. First, I briefly recall whether patterns, like business cycles, are indeed found, and can be modeled within a statistical physics, or econophysics,  framework.  I point to a simulation model for describing  such so called business cycles, under exo- and endo-genous conditions I  discuss self-organized and provoked crashes and their predictions. I  emphasize the role of an often forgotten ingredient: the time delay in the information flow. I wonder about the information content of  financial data, its mis-interpretation and market manipulation. 

\end{abstract}


\maketitle

\section{Introduction  } \label{intro}

The time is ripe for discussing recent economic  and financial crises timing. Are they recurrent? Is there any universality, i.e.  "universal laws", in the physics strict concept sense to describe such events? Are the present  financial fluctuations, similar or not, to the spectacular financial crashes of stock markets of past years? How quantitative can be a  warning?  Can one relate economic "theoretical models" to  perceived features? Is there any manipulation of  "the" economy and financial markets?  Are there basic faults, voluntary or not, in markets and economy policies? (The latter question reminds of some argument  which at some time  claimed that  the misuse of the erroneous
  Black-Scholes equation was the cause of the financial crisis. But is one equation usage only to blame for the crises that  existed before the equation, - and for those which have followed?) 

There always exists the hope that mathematics can supply tools and methods to predict when events  happen, but first one needs to understand the processes that produce them for elaborating realistic evolution equations, - and subsequent forecasting.  Thus, economic considerations  could be usefully combined with  physics models of nature. Working by analogy,  econophysicists,  {\it mutatis mutandis},  introduce economic  rules and interactions  as   basic ingredients in concise models, like replacing "spins"  by "agents", to mimic social and economic phenomena. Yet, even if econophysicists do  control their models, of course, it is known that complex systems  generate unpredictable outcomes.  However,  they present a series of general features that we can study. Whence,  I will briefly outline a few considerations on business cycles and on spectacular features like crashes. I will insist upon subtle ingredients which are much part of present economic life, though they existed also much before, i.e. information flow, market manipulations by active agents and by media or politicians.

\section{So called cycles} \label{cycles}

Let us start with some optimistic thought leading to things not necessarily much discussed, though  supposedly of common knowledge :  "things will go better".  This is equivalent to the classical statement that "the   darkest hour of all  is the hour before day"; equivalently,  in french,  "apr\`es la pluie, le beau temps".   A mere meteorological evidence, of no  quantitative forecasting value. Necessarily, indeed, after a recession, will come prosperity.

To improve such views, one has imagined to quantify "business cycles", i.e.,  merely the durations of recession and prosperity time intervals and frequencies.  Since 1930 or so, observations of  "cycles" had led Kalecki \cite{Kalecki} to write about a macrodynamic theory of business cycles.  Let Gabsich and Zarnowitz books  or Aoki contribution \cite{others1,macro,aoki}   be  also mentioned, but there are many other  relevant works and many controversies \cite{wikiABCT} to be but which have been much discussed.

 However, it is extremely hard for a physicist to accept the existence of such features as    being a cycle, when there are about a little bit more than 100, or even 200,  data points, and when the deduced cycle period is about 35, or 55,  years, - with large, not to say huge, error bars on this "period" value. Business cycles are still theoretical concepts. NO theory can predict the time interval size of a recession or  prosperity;  still less its day, or even month,  of beginning or end.

\subsection{Duration distributions}
Yet, following recent practical use of statistical physics concepts in financial matters,  Ormerod and Mounfield  \cite{ormer}  analyzed  Gross Domestic Product  (GDP) data of 17 leading capitalist economies from 1870 to 1994 and concluded that the $frequency$ of the $duration$ of  $recessions$ is consistent with a power law.    The data  analysis was challenged by Wright  \cite{wright} who concluded that the data was consistent with an exponential (Boltzmann-Gibbs) law.  The controversy might stems, as already hinted here above, from the number of data points used in measuring the shortest time intervals for a recession.   

Thus, do the recession duration distribution follows a power law or an exponential law \cite{sanglierauslooscycles}? An extended set of GDP data was  examined in  \cite{sanglierauslooscycles}. It was   concluded that a power law led to a better fit than an exponential law over the extended data set.  It seemed of interest also  to   examine  $prosperity$  $duration$.  Interestingly,  the distribution  was found to follow a power law \cite{sanglierauslooscycles}.  

Thus, the arguments of Ormerod and Mounfield  about the duration frequency of recessions might be recalled, if not fully approved or considered to be exhaustive: the authors proposed that economic management often prevents recessions lasting for more than 1 year, but if they do last longer, then subjective expectations of growth become depressed and recessions may then occur on all scales of duration, resulting in a power law. They proposed, therefore, that the distribution of such  patterns is not determined by a common set of causal factors for all durations, but instead there is a Ôbreakdown of scalingÕ for recessions with short duration. Question:  can these ideas be extended to the cases of prosperity epochs?

Note that often there is some perception difference between the psychological or even visual data filtering based on the slowest trend, - mathematically, a large window-moving average, and the actual  results from statistical analyses, looking at somewhat high  frequency data.

\subsection{Pulse destabilization}
Considering that the economy is basically a bistable (recession/prosperity) system, it was suggested   \cite{ormer} that   a characteristic (de)stabilization  model be considered. There are many possible theoretical  ones, but each should lead to a characteristic (de)stabilization  time, in order to be of conceptual value.  

However, let it be emphasized that from an economic point of view this type of pulse time values should not be confused with the apparent periodicity (or better ÒperiodicitiesÓ) of business cycles 
\cite{Kalecki,others1,macro,aoki},  closely related to what can be called  a relaxation time,  in physics. 
 A pulse (force) duration much differs from the time during which the force strength will have some measurable  effect   \cite{PNAS75.78.4633-7-Montroll-socialforces,AUsloosACS}.  Note that the pulse application anticipates an up or down turn, or sometimes an acceleration or a deceleration \cite{MAladek,4coher}. By "how much?",  is a pertinent question. It depends on the pulse strength and its duration, and moreover on what state it is applied, - since the system is always in a dynamic equilibrium!

Marchetti \cite{Marchetti} had  observed cycles and pulses behavior of economic indicators, leading to "business
cycle" features, after correlating them to
 physical and other patterns of social behavior, rather  than to 
 money indicators, - with a period of about 55 years for at least two centuries.   The patterns can be expressed in close mathematical form permitting quantitative forecasts to be made. Along such a psychological line of expectations for economy periods, Hohnisch et al. \cite{Hohnisch}  discussed the ups and downs of opinions about recessions and prosperity within a Glauber dynamics evolution frame \cite{Glauberdyn,GlauberdynCasti,GlauberdynStauf} of a Blume-Capel model \cite{BlumeCapel,BlumeCapelMFA,BlumeCapel3crit}, which in brief describes ''spin 1'' particles. Their computer simulations show that opinions drastically undergo changes from one equilibrium to the other, both having rough but small fluctuations, - as in stochastically resonant-like systems.
These opinion swings are random events like radioactive decay, and thus presumably exponentially distributed. 
  
  The (observed) power laws \cite{ormer,sanglierauslooscycles} thus indicate some coupling between successive swings, which is missing in the simple computer model of Hohnisch et al. (Similarly, the simple random walk model of Bachelier 1900 gives a Gaussian or log-normal distribution of the price fluctuations, and the actually observed power law distribution needs a non-random explanation, like psychological herding etc.). Some  further discussion of the meaning of exponential versus power law decays thus would be appropriate, but is outside the  scope of the present comments. One attempt, seen as an inverse problem, should be to invent stochastic time series with appropriate structures as in  \cite{cr,FPMAGR}.

Note that there are asymmetries in the  prosperity and recession sections, with different statistical characteristics. There are also more subsections in the  prosperity than in the recession time intervals. This asymmetry  much complicates the analysis.  Is it time to suggest further work ion the matter? One could make a Detrended Fluctuation Analysis \cite{DFA,DFA1,DFAma}, or use other related techniques  \cite{[23],[35],luxma,MAKI402CPC}.  This would allow to  take  into account that overall the Western world had economic growth in the last half century, characterized by a time-averaged growth rate G per year or per quarter. Then  a  time interval with actual growth above (or below) G could be counted as prosperity (or as a recession). In this way, the distribution  and its analysis would be perhaps better suited to finer details, as the fluctuations and their correlations. Quite recently, I showed thatsuch fluctuations in a G averaged over  finite time intervals can reveal trends and superposed humps, in some financial data, - surprisingly with regular evolutions of the time interval durations \cite{MAantoinist1}.

 In concluding this section, I wish to  indicate that such successions of recessions and  prosperity periods, thus so called business cycles, have pushed several of us into  a  renewed deal of attention, from analytical and simulation points of view \cite{aoki,ACPbali1,JMMAlogi,JMMAlg}.

\subsection{ACP  model: a Reactive Lattice Gas} \label{ACPmodel}
In particular, in the so called ACP model \cite{ACPbali1}, one simplifies economic rules to simulate the behaviour of economic systems in changing conditions, e.g. political changes or destruction of economy barriers  \cite{ACPbali1,ACP3r,ACP3BS}. This is similar to a Boltzmann  approach in which 
the gas particles are  called companies, but this can also be only a name
for any economic variable \cite{JMMAlg}. The company ''degree of
freedom'' (its particle efficiency  or fitness $f$) can be coupled to an external field $F$, describing the political or /and economic environment.
Beside the field, there is some selection pressure $sel$, which acts as an
inverse temperature in a thermodynamic  sense \cite{ACPbali1}.

Economy entities were initially randomly placed on a square symmetry lattice.   We have also considered a world divided into three regions characterized by different taxes, or  economic rules \cite{ACP3r}.
Three possible events   determine the evolution of  the system:
\begin{itemize}\item 
  a company $i$  survives with the probability $p_i = e^{-sel\; \|f_ i -F\|}$; 
  \item  each company can move on the lattice (horizontally or vertically) one lattice step at a time, if a free space is available; 
  \item if companies "meet" they can set a new firm or  can merge; the decision is taken through the strategy parameter $b$, which is the probability of merging. Since the creation of company is a complementary event the settlement probability is  $1 - b$.
  \end{itemize}
  
 The $f$ value of the new firm was considered
to be obtained according to various types of memories depending on the $f$ of the
company parents. Results of variants of such a model indicate the possibility of finding business cycles  \cite{JMMAlogi,ACP3BS}. Moreover, in  order to investigate the role of the time delay in information flow, the model was adapted to the Ising model of financial markets \cite{sornette} by introducing interactions between the stock market and the economy as well as the time delay in the information transfer about the economy state \cite{JMMAvd,JMMAdelay}. Further work should lead to determine some (de)stabilization  time and to more quantitative results for mapping economic reality to the model, i.e. to better describe in an economic way, how the parameters $F$, $f_i$ and $sel$  can be deduced and more precisely defined, from considerations on "manufacturing" and "agricultural" sectors, "production factors", like "entrepreneurs" and "unskilled labour" , as, e.g.,  in the Footloose Capital model \cite{FCM} and its extensions.

\section{Financial crashes}\label{logP}

\subsection{Endogenous vs. Exogenous crashes}

Going to the argument of predictability, one has to touch upon the notion of self-organization \cite{Bak} and of external field influence.  In fact, the lack of universality of crises is likely to be found in differences between endogenous and exogenous origins of interactions between agents under their environment constraints, modulated by media information and other agents, inducing different reactions \cite{sornette0210509v1,sh,sornette0412026v1}. 
There is much quantitative difference between endo and exo-genous  effects. E.g., a simple concise case which I studied is that of sales in which one can show different relaxation times \cite{sales}.
Endogenous effects  are partially due to psychological impacts of traders and financial manipulators. Not only.  Of course, they went to different economic schools. If they went to school, of course they learned different theories,  and their gurus forced them into being devoted followers or original contrarians.  
 Exogenous effects  emphasize similarity in reactions. Of course, at the microscopic level, people will react in different ways, to an earthquake or a tsunami, but globally the same herding behavior is found. We are educated, told, informed to copy others;  yet, there are sometimes contrarians, who  thus make "live" evolving in a stochastic way. The same in economy and finance. Everyone necessarily relies on information for behaving. 
 
 Nowadays those informations are much quantified; I do not  claim that all computers in all trading firms are based on the same codes, but they rely on similar  informations, and regularly accepted theoretical rules. There are variations, but the sell-buy mechanism is intrinsically accepted as the basic ingredient of  the financial system, and  the instantaneous balance between trades  is a strong hypothesis. Of course, there is a relative importance  of agents, more than of news, in the evolution  of the markets, because one follows "big shots" behaviors. Note that leaders  can also manipulate in order to win, - see below.  
 
 Computers  are supposed to $quasi$ react in the same way. However, 
in most model, time delay in the information flow is $very$ often missing. Yet, it is one cause of
 automatic amplification effects, which are crucial at crises times.
Reaction synchronization in crisis or in non-crisis "periods" differs for both exogenous and endogenous crash cases. Fast oscillations are superimposed on low evolutions, or the other way around, as easily understood, but with markedly different time scales.  To remain in a sort of "equilibrium dynamics framework model" of markets, and use physics vocabulary, let us write :  Similarly, to phase transitions in thermodynamics, one can imagine that a self-organized crash should be independent of  external fields, and  be typically deterministic,  yet containing some uncertainty. However, manipulations  can be thought of as external fields.

That is why, when the endogenous multiplicative effect prevails over exogenous causes that one can predict crashes. This was markedly obvious when the Oct. 27, 1997 crash was predicted several months in advance \cite{TTcrash1,cash1,TTcrash2,cash2,how},  the warning signals being extremely similar to the Oct. 19, 1987 crash analyzed in \cite{Feigenbaum96,SornetteAJBouchaudJFI96}  through the log-periodic functional form
\begin{equation}\label{logPm}
y(t) = A + B {\left( {t_c-t \over t_c} \right)}^{-m} \left[ 1 + C \left(
\omega \ln{\left( {t_c-t \over t_c} \right)} + \phi \right) \right]  
  \hskip 0.3cm  (t < t_c).
\end{equation} 
where $t_c$ is the crash-time or bubble rupture point, - the other symbols being
parameters,  which can even, by analogy, receive some reasonable interpretation \cite{bigtokyo}.  The same oscillatory pattern before a divergence-like behavior also occurs for the 1929 crash(es). This is as if the market is a physical system at an equilibrium second order phase transition but evolving  under non-equilibrium conditions \cite{cp,MAapl43,MAriste}. 

No consensus exists about the precise (economic or financial) causes, of such types of crashes, but the fits to the financial indices fluctuations are remarkable \cite{grech1,grech2,DStQF02}. A longer, but still incomplete,  list of references can be found in \cite{DSornPR03,DSornbookcrash,XZhDSorn03,DSornXZh06,DrozdzEPJB99,kwapiendrozdzPR12}. This suggests that for  several crashes some underlying "structural" mechanism exists \cite{JR1.99.5CrashSornette,DSIDSornPR98}. It is conjectured that this holds for self-organising-like systems. In the case of exogenous  effects,  unexpected or under manipulation, the universal law, Eq.(1), or its variant \cite{how}, has $not$ to hold, - see e.g. what has  been recently observed and is outlined here below.

\subsection{Cerdeira index}\label{cerdeira}

Obviously, the question  still remains on whether one can be identifying bubbles and subsequent  financial crises  in real time. To do so, see \cite{redelicoproto} and  \cite{cerdeiracrises}. E.g., a measure, called  the area variation rate (AVR) index,  has  been introduced  to study the behavior of  financial  time series in \cite{cerdeiracrises}.  The index is defined as the ''normalized  (free) energy'' variation, deduced from a sort of ''specific heat'' integration over  "time windows", as in  a multifractal approach \cite{luxma,MAKI402CPC}. This method  permits to distinguish   and identify rare events, like major  financial crises, as the  (Oct. 24) 1929 Black Thursday and  the (Oct. 19)  1987 Black Monday market crashes  and the   2008 Subprime crisis.  The analysis has forecasting capabilities, as it has been demonstrated when  applying the method  to the  2011 market fluctuations. From these results, it appears in fact  that the 2011  crisis   seems of a different nature from the other three here above mentioned \cite{cerdeiracrises}. Should  I dare conjecturing that, beside market manipulation, - see below,  a structural cause might be the value of the information flow time delay at the various (= different) crisis epochs?

  \section{Market manipulation}\label{Jmanip}

 \subsection{Market manipulation by agents }\label{agentmanip}

Some   market manipulation in the financial crisis has been recently evidenced  \cite{necsimanipul}.  

An analysis of financial industry data,  Citigroup's stock, has shown an anomalous increase in borrowed shares at some time, - the selling of which  being a large fraction of the usual total trading volume.  Such a selling of borrowed shares could not be explained by news events, as there was no corresponding increase in selling by share owners. However a similar number of shares, as the borrowed ones,  was returned on a single day six days later.   The magnitude and coincidence of borrowing and returning of shares (of Citigroup's) is evidence of a concerted effort to drive down Citigroup's stock price and achieve a profit.

On the other hand, I have  come across "obvious" insider trading in one case, based on a DFA data analysis ,  -  I confess it, leading to some personal financial loss. But can a judge rely on an econophysicist  suprising finding?, - and could such an analysis  be published, as a scientific fact? 
 
 It is accepted or admitted as ''common knowledge''  that markets are not sufficiently transparent to reveal or prevent   market manipulation events, like a "bear raid,"  or insider trading.  Note, however, that an appropriate regulation, repealed by the Securities and Exchange Commission in July 2007,  the so called "uptick rule", was designed to prevent such market manipulations and even promote so called stability.  It was in force from 1938 as a key part of the government response to the 1929 market crash and its aftermath. Not any more. It might be reinstated,  see $http://en.wikipedia.org/wiki/Uptick_-rule$.

\subsection{ Market manipulation by media}\label{JS}
The following stems from a private communication by D. Stauffer \cite{Janssen-Staufferobservation}. 
 
Media     present Germany as debt-free prigs, and even 
described  Germany as  the model of   debt delimitation. In reality,   according to  the
Statistics Office of the European Commission, Eurostat, more  than two thirds of the 27 states of  the European Union have less national debt (with respect to the Gross Domestic Product) than Germany.

In the  spring of 2010, Federal Chancellor Angela Merkel suggested in "Bild am SonntagÓ   that   permanent  "Schuldens\" under" should loose their national right to vote in Europe;
neither the chancellor nor  "Bild'' added that Germany should be part of them.   Why? ... because they did not know the data, or because they knew that the German people did not know it? The more so, since
former chancellor Helmut Schmidt   had written  in "Die Zeit'', in 2011, that only one country fulfilled the two Maastricht criteria
(60 percent of debts, and 3 percent yearly new indebtedness = deficit).  In fact,   5 European Union states do so: Luxembourg, Sweden, Finland, Estonia, Denmark and (only for 2011) Bulgaria. According to Eurostat, in 2011 (provisionally), in the European Union, the highest national debt ratios were: Greece, Italy, Ireland, Portugal, Belgium, France, Great Britain, Germany, and Hungary.
  
 When,  in April 2011, Eurostat  released the provisional statistics for  2010,  only    "Die Welt"   among the most  important   (german) newspapers published them.
   "Der  Spiegel" brought the 2011  crisis  as a cover story, but  wrote more about  Greece than about Germany. {\it Finally}, in the fall of 2011,  newspapers reported in detail  an alleged false estimation of the German national debt,  as being  "small",  which was in reality around 55 billions euro. 

Of course, media do report dissenting voices:  the  Luxembourg Minister-Pr\"asident and Chairman of the Euro group, Karl Juncker claimed that Germany has more state debts in percent of the GDP than Spain. 
 Moreover,   in one
interview with  "Der Spiegel" (p. 71, in the 07/2012 issue) financier George Soros said that:  ''Germany was one of the states, which freely offended   the (Maastricht criteria)''. However, media are prone to moderate such remarks. E.g.  ''Der Spiegel'' countered:  Germany has $occasionally$ overdone the three-percent limit in the budgetary deficit.

All this to show how on a specific, but highly political case, media can  (by omission  or  for some other reason which cannot be written here)  induce a false vision of a (financial, here)  situation.

\section{CONCLUSION  } \label{concl}

Events such as business cycles, market crashes and economic crises are ubiquitous 
 and generate anxiety and panic difficult to control. Explaining, preventing or forecasting  them has been the chimera for  many economists, politicians and even scientists who have devoted much of their work to the matter. 
  To describe these events, one needs to take into account an enormous number of ... unknown variables. However,  this does not impede anyone from finding  temporal correlations or patterns in the interesting time series.
  
   
  Moreover physics intuition and {\it ad hoc}  statistical models may give some hint on the main highlights because of the pertinent quantification of so called  stylized facts. My impression, nevertheless,  is that sound scientific work has to fight, and will have often to fight, against  (scientific or not) tradition and political constraints.

\vskip 0.5truecm

 {\bf Acknowledgements} \vskip 0.5truecm

The author thanks Dr. R. Cerqueti for his longtime friendship, but recognizes that  a challenge, such as the one leading to the present scientific contribution,  was not expected.

\end{document}